\documentclass[11pt]{article}
\usepackage{graphicx}

\begin{document}

\def\xslash#1{{\rlap{$#1$}/}}
\def \p {\partial}
\def \dd {\psi_{u\bar dg}}
\def \ddp {\psi_{u\bar dgg}}
\def \pq {\psi_{u\bar d\bar uu}}
\def \jpsi {J/\psi}
\def \psip {\psi^\prime}
\def \to {\rightarrow}
\def \lrto{\leftrightarrow} 
\def\bfsig{\mbox{\boldmath$\sigma$}}
\def\DT{\mbox{\boldmath$\Delta_T $}}
\def\xit{\mbox{\boldmath$\xi_\perp $}}
\def \jpsi {J/\psi}
\def\bfej{\mbox{\boldmath$\varepsilon$}}
\def \t {\tilde}
\def\epn {\varepsilon}
\def \up {\uparrow}
\def \dn {\downarrow}
\def \da {\dagger}
\def \pn3 {\phi_{u\bar d g}}

\def \p4n {\phi_{u\bar d gg}}

\def \bx {\bar x}
\def \by {\bar y}

\begin{center} 
{\Large\bf   Parton Interpretation and Twist-4 Parton Distributions }
\par\vskip20pt
 J.P. Ma$^{1,2,3}$,  Z.Y. Pang$^{1,2}$ and G.P. Zhang$^{4}$    \\
{\small {\it
$^1$ CAS Key Laboratory of Theoretical Physics, Institute of Theoretical Physics, P.O. Box 2735, Chinese Academy of Sciences, Beijing 100190, China\\
$^2$ School of Physical Sciences, University of Chinese Academy of Sciences, Beijing 100049, China\\
$^3$ School of Physics and Center for High-Energy Physics, Peking University, Beijing 100871, China\\
$^4$ Department of Physics, Yunnan University, Kunming, Yunnan 650091, China}} \\
\end{center}

\vskip 1cm
\begin{abstract}
 Through explicit examples we show that twist-4 parton distributions have no parton interpretation in the sense that parton or partons inside a hadron can carry the momentum fraction $x$ of the hadron with $x >1$ or $x<-1$.  The studied twist-4 parton distributions of collinear factorization are power-divergent for $\vert x\vert >1$.  The corresponding transverse momentum dependent parton distributions have 
finite contributions for $\vert x\vert >1$ and they have hence  no parton interpretation. 
Parton distributions are defined with products of field operators. In the case of twist-4, different orderings give different results. The differences are determined by commutation relations of field operators. This has been 
explicitly checked in the studied case. 
The implications of our results are discussed.

 \end{abstract}      
\vskip 5mm
\noindent

\par\vskip20pt
\noindent

Cross-sections of high energy scattering with a large momentum transfer can be predicted at leading power with QCD 
factorization, in which twist-2 parton distributions are convoluted with perturbative coefficient functions. 
These parton distributions describe the motion of partons inside a hadron and hence contain information 
about the inner structure of hadrons. Beyond the leading power, higher-twist parton distributions appear 
in the factorization. With experimental progresses, it is now possible to study higher-twist effects. 
In order to consistently extract from experiment higher-twist parton distributions their properties 
need to be studied in theory. 

In high energy scattering, an initial  hadron can be viewed as a bunch of quarks and gluons, i.e., partons. 
In a scattering with large momentum transfers, a parton or partons from the hadron take part in the hard scattering. 
A basic property of twist-2 parton distributions of a hadron is that the distributions have only the support of $x \in [0,1]$, 
i.e., the parton can not have the momentum larger than that of the hadron, or the parton in the initial state does not absorb energy from hard scattering.  This property ensures that the twist-2 parton distribution has the so-called parton interpretation.  It has been explicitly shown in \cite{T4Ja}  that twist-2- and twist-3 parton distributions have such a property, i.e., they have the parton interpretations.
Twist-2 
generalized parton distributions also have  parton interpretation as shown in  \cite{DiGo}. 
 
\par 
It is unclear if twist-4 parton distributions have a parton interpretation, although it is believed that to be the case based on \cite{T4Ja}. In this letter we show that twist-4 parton distributions are nonzero for $x>1$ or $x<0$, i.e, they do not have the parton interpretation. 
In defining parton distributions products of field operators are used. 
For twist-2 and twist-3 it is irrelevant whether the products are time-ordered or not. From this fact one can show the property which leads to  the parton interpretation.  
In the case of twist-4, there is a difference in the distributions defined with or without time-ordering as we will show. The difference can also be calculated from canonical commutation relations.

It will be helpful for our purpose to discuss the twist-2 quark distribution of a single quark  at one-loop to show 
how the parton interpretation appears.  Usually in literature one has only considered the contribution with $x>0$. 
This contribution is in fact nonzero for $x<0$. This is unphysical at the considered order. 
However, there is in addition an unmentioned contribution with $x<0$, which has a difficulty to be calculated.
In \cite{JCBook} it is shown that the sum of the two contributions is zero in a Yukawa theory. 
 We will calculate the additional contribution explicitly and find that the distribution at one-loop is zero for $x<0$.  With the method used here, one can calculate the studied twist-4 parton distributions straightforwardly.

We use the  light-cone coordinate system. In this system a
vector $a^\mu$ is expressed as $a^\mu = (a^+, a^-, \vec a_\perp) =
((a^0+a^3)/\sqrt{2}, (a^0-a^3)/\sqrt{2}, a^1, a^2)$ and $a_\perp^2
=(a^1)^2+(a^2)^2$. We introduce two vectors in the system: $n^\mu=(0,1,0,0)$ and $l^\mu =(1,0,0,0)$. 
Throughout our work we use the light-cone gauge $n\cdot G=G^+=0$ with $G^\mu$ as the gluon field.  

The twist-2 parton distribution of an unpolarized hadron with the momentum 
$P^\mu =(P^+, P^-, 0,0)$ is defined as: 
\begin{equation} 
 f_q (x) =  \int\frac{d\lambda}{4\pi} e^{-i\lambda x P^+} \langle P \vert \bar \psi (\lambda n) \gamma^+ \psi(0) \vert P\rangle. 
 \label{FQ} 
\end{equation} 
It is noted that in the definition the product of the two quark fields is not time-ordered. One can insert 
the unit operator between the two quark fields
\begin{equation} 
     1 =\sum_X  \vert X\rangle \langle X\vert 
\label{CPR}      
\end{equation} 
to show that $f_q$ is zero for $x>1$. Therefore, $f_q$ has the parton interpretation that the parton carries the momentum $xP^+$ as a part of the hadron momentum.  The momentum fraction is $x$.         
 The parton interpretation also requires that  the parton entering hard scattering must have  $x>0$. 
In the definition, one can also take the operator product as time-ordered, because of that  it can be shown in \cite{T4Ja,T2PDF}  that $f_q$  defined with time-ordered 
product of field operators  is the same as that in Eq.(\ref{FQ}). This will also be discussed later. 

\par\vskip5pt
\begin{figure}[hbt]
	\begin{center}
		\includegraphics[width=12cm]{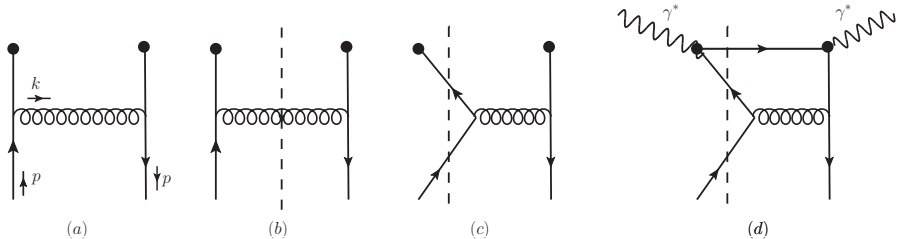}
	\end{center}
	\caption{(a): Uncut diagram for one-loop correction. (b)and (c): Cut diagrams for one-loop real correction.  (d): A diagram of DIS with a quark target at one-loop with $x<0$, its contribution is zero because
	of  conservation of energy.   }
	\label{TreeL}
\end{figure}
  
It is interesting to calculate $f_q$ of a single quark state.  At tree-level, it is easy to obtain 
$f_q (x) = \delta (1-x)$.  The virtual correction at one-loop is proportional to $\delta(1-x)$. 
The problem mentioned with $x<0$ appears in the real correction at one-loop.  If $f_q$ is defined 
with time-ordered 
product of field operators, the correction is given by the uncut diagram in Fig.1a. After working out the color factor, the correction is: 
\begin{eqnarray} 
f_q (x) \biggr\vert_{1a} = g_s^2 C_F\frac{1}{4}  \int\frac{dk^- d^2k_\perp }{(2\pi)^4}      \frac{{\rm tr} \left [ \gamma^+( \gamma\cdot (p-k)+m)  \gamma^\mu (\gamma\cdot p +m)\gamma^\nu (\gamma\cdot (p-k)+m)  \right ] } { ( (p-k)^2-m^2 + i\varepsilon ) ( (p-k)^2 -m^2 + i\varepsilon ) (k^2 +i\varepsilon) } 
     (-i)    {\mathcal N}_{\mu\nu} (k), 
\end{eqnarray} 
where $k^+$ is fixed as $k^+=(1-x) p^+$ and  $m$ is the quark mass. 
 ${\mathcal N}_{\mu\nu} (k)$ is given by the numerator of the gluon propagator in the light-cone gauge: 
\begin{eqnarray}          
    -i {\mathcal N}_{\mu\nu} (k) =-i       \biggr ( g_{\mu\nu} - \frac{n_\nu k_\mu + n_\mu k_\nu}{n\cdot k} \biggr ). 
\end{eqnarray}            
To perform the integral of $k^-$, we search for poles in the complex $k^-$-plane.  It is straightforward to find that in the case of $1>x>0$ there is a single pole from 
 the denominator $k^2+i\varepsilon$ in the lower half-plane and a double pole from the denominator $( (p-k)^2 -m^2+ i\varepsilon )^2$ in the upper half-plane, as shown in Fig.2.  In other cases, all poles are either in the upper- or the lower  half-plane.  We can take a closed contour encircling one pole as shown in Fig.2 to perform the integration. In other cases, we can take the same closed contour in the lower- or upper 
 $k^-$-plane so that there is no pole inside the contour.   
 In the case of $f_q$ the contour integration along the semicircle is zero in the limit when its radius becomes 
 infinity large.  We have then the result:
 \begin{equation} 
  f_q (x) \biggr\vert_{1a} = 2 \alpha_s C_F   \theta (x) \theta(1-x) \int \frac{d^2 k_\perp}{(2\pi)^2} \biggr [ \frac{1+x^2}{1-x} \frac{1}{k_\perp^2 +(1-x)^2 m^2}  + \frac{m^2 x}{ (k_\perp^2 +(1-x)^2 m^2)^2 } \biggr ] .  
 \label{FQ1A}  
\end{equation}   
When we perform the integration over $k_\perp$,  the result contains an U.V. pole in $\epsilon=4-d$ which will be subtracted, a collinear divergent contribution in the limit $m\to 0$, and a finite part.  We leave the integration here  because it does not affect our discussion. It is interesting to note that the result from the uncut diagram, or in the case 
that $f_q$ is defined with time-ordered product, satisfies the requirements of the parton interpretation, because of that $f_q$ is zero for $x>1$ or $x<0$.

\par\vskip5pt
\begin{figure}[hbt]
	\begin{center}
		\includegraphics[width=4cm]{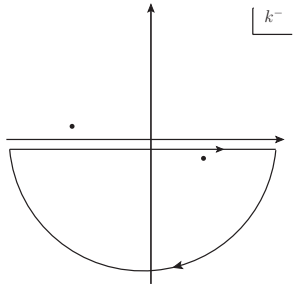}
	\end{center}
	\caption{ The pole position in the complex $k^-$-plane in the case $1>x>0$. The black dots are poles.     }
	\label{TreeL}
\end{figure} 

In the case $f_q$ is defined as in Eq.(\ref{FQ}) where the product of operator is without time-ordering, 
the real correction is from Fig.1b with one gluon in the intermediate state. 
This is obtained by using partonic completeness relation like that Eq.(\ref{CPR}) at one-loop level. 
It is noted that using partonic completeness relation in DIS instead of hadronic one can have certain differences, a discussion about this can be found in \cite{DAL}. In this letter we will calculate parton distributions with partonic states and use the partonic completeness relation. 
The contribution from Fig.1b is:
\begin{eqnarray} 
f_q (x) \biggr\vert_{1b} &=& g_s^2 C_F\frac{1}{4}  \int\frac{dk^- d^2k_\perp }{(2\pi)^4}     (-2\pi i ) \delta (k^2) 
     (-i)    {\mathcal N}_{\mu\nu} (k)
\nonumber\\
  && \cdot  \frac{{\rm tr} \left [ \gamma^+( \gamma\cdot (p-k)+m)  \gamma^\mu (\gamma\cdot p +m)  \gamma^\nu (\gamma\cdot (p-k)+m)  \right ] } { ( (p-k)^2-m^2 + i\varepsilon ) ( (p-k)^2 -m^2 - i\varepsilon )  } ,     
\end{eqnarray} 
with $k^+ =(1-x)p^+$.  It is straightforward to obtain 
the result:
 \begin{equation} 
  f_q (x) \biggr\vert_{1b} = 2 \alpha_s C_F   \theta(1-x)   \int \frac{d^2 k_\perp}{(2\pi)^2} \biggr [ \frac{1+x^2}{1-x} \frac{1}{k_\perp^2 +(1-x)^2 m^2}  + \frac{m^2 x}{ (k_\perp^2 +(1-x)^2 m^2)^2 } \biggr ] .  
\end{equation}    
Because the gluon in Fig.1b is real, one always has $k^+>0$. This gives the factor $\theta(1-x)$ in the above. 
We note here that this contribution is in fact not zero for $x<0$. 
If the real correction is only from Fig.1b, then $f_q(x)$ defined in Eq.(\ref{FQ}) does not have the parton interpretation and it is different than $f_q(x)$ defined with time-ordered product of operators, i.e., different 
than that from Fig.1a in Eq.(\ref{FQ1A}). 

If we consider the factorization of DIS at one-loop, the contribution from Fig.1b with $x>0$ corresponds to the physical process $\gamma^* + q\to q$, where the initial quark 
carries the momentum fraction $x$, while  the contribution with $x<0$ to $f_q$ corresponds to the process $\gamma^* \to q \bar q$ as shown in Fig.1d, where the antiquark crossing the cut   
carries the momentum fraction $-x$. 
The later is not allowed 
because of that the virtuality of the photon is negative.  $f_q$ can be nonzero for $x<0$ only beyond one-loop.  At one-loop there are in fact additional contributions from Fig.1c and its complex conjugated diagram. 

The contribution from Fig.1c is:
\begin{eqnarray} 
f_q (x) \biggr\vert_{1c} &=&  g_s^2 C_F\frac{1}{4} \theta (-x)  \int\frac{dk^- d^2k_\perp }{(2\pi)^4}   \frac{-2\pi i \delta ((p-k)^2-m^2)}{(p-k)^2-m^2}     
     (-i)    {\mathcal N}_{\mu\nu} (k)
\nonumber\\     
  && \cdot \frac{{\rm tr} \left [ \gamma^+( \gamma\cdot (p-k)+m)  \gamma^\mu (\gamma\cdot p+m)  \gamma^\nu (\gamma\cdot (p-k)+m)  \right ] } {k^2 - i\varepsilon }   .  
\end{eqnarray} 
At first look, this contribution has the difficulty to be calculated because of the divergent factor $\delta ( (p-k)^2-m^2) /( (p-k)^2-m^2)$.  The factor appears because we have in the both sides of Fig.1c the same quark propagators. The one 
in the left side is cut. 
 This divergent factor should be regularized by a certain limit. 
 In \cite{JCBook} a similar case with the quark distribution defined  in Yukawa theory has been analyzed. By taking different momenta of quark states, it has been shown that the contributions  with $x<0$ are cancelled\cite{JCBook}. 
 In this letter we use a slightly different method to calculate the contribution from Fig.1c and its complex conjugated diagram.  We show in the below with 
a simplified case that the contribution from Fig.1c and its complex conjugated diagram is the 
same as the double pole contribution from Fig.1a mentioned after Eq.(4).    

We consider the following expression as the mentioned simplified case:
\begin{equation} 
   I =  \int_{-\infty}^{\infty}  dz \biggr [  \frac{2\pi  \delta (z-a)} {z-a} f (z) + \frac{2\pi  \delta (z-a)} {z-a} f(z) \biggr ] , 
\label{DI} 
\end{equation} 
where $f(z)$ is real when $z$ is real. This expression is divergent and should be regularized. 
We take $I$ as the limit, in which $a$ and $b$ are real:    
\begin{equation} 
    I  =\lim_{b\to a}  \int_{-\infty}^{\infty} dz 
    \biggr [ \frac{2\pi  \delta (z-b)} {z-a} f (z) + \frac{2\pi  \delta (z-a)} {z-b} f(z) \biggr ].  
\label{EQ53}       
\end{equation} 
In the case with Fig.1c, we can regularize the divergent factor by giving the incoming- and out-going quark 
different momenta so that the two quark propagators in the both sides of Fig.1c are different. 
After the regularization, the integral of $k^-$ is similar to the simplified case here, where one can identify that  the first- and second term in Eq.(\ref{EQ53}) 
correspond to the contribution of Fig.1c and that of it complex conjugated diagram, respectively. 
 Now we can express the integral  $I$  as an integral along a close contour $c$ in the complex $z$-plane.  The close contour $c$ only encircles 
 the two poles at $z=a$ and $z=b$ anti-clockwise.   Taking the limit $b\to a$ we have: 
\begin{eqnarray} 
I   =  \frac{1}{i}  \lim_{b\to a} 
     \oint_{c} dz \frac{1} {(z-a+i\varepsilon) (z -b+i\varepsilon)} f (z)  = \frac{1}{i} \oint_c dz \frac{1}{(z-a+i\varepsilon)^2} f(z).    
\end{eqnarray} 
After the limit, there is only a double pole at $z=a$ inside the close contour $c$. Therefore, the expression 
$I$ in Eq.(\ref{DI})  is the contribution from the double pole.       
\par 
Now we can evaluate the contribution from Fig.1c and its complex conjugated one. As discussed in the above, the contribution from the two diagrams is the contribution from the double pole in the complex $k^-$-plane. We have:
 \begin{equation} 
  f_q (x) \biggr\vert_{1c+c.c.} =  - 2 \alpha_s C_F   \theta(-x) \int \frac{d^2 k_\perp}{(2\pi)^2} \biggr [ \frac{1+x^2}{1-x} \frac{1}{k_\perp^2 +(1-x)^2 m^2}  + \frac{m^2 x}{ (k_\perp^2 +(1-x)^2 m^2)^2 } \biggr ] .\end{equation}   
Adding this contribution, we obtain that $f_q(x)$ is zero for $x>1$ and $x<0$. This agrees with the result from the uncut diagram and satisfies the requirements of parton interpretation at the considered order.   
We notice here that the method used here to deal with the contribution from Fig.1c can not be used in the case that the contribution is singular for $x\to 0$, like the case of the twist-3 parton distribution $e(x)$ which has a singular contribution proportional to $\delta (x)$ at $x=0$ in \cite{BuKo, EX}. As mentioned in the introduction part, as a twist-3 parton distribution $e(x)$ has the parton interpretation.  An one-loop analysis in \cite{EX} has shown this explicitly. 
\par 
Now we turn to cases of twist-4 parton distributions. We first discuss  one of  the twist-4 parton distributions
appearing in the parameterizing twist-4 effects in DIS. The analysis of twist-4 effects in DIS can be found in  \cite{JaSo, EFP, JWQ}. We take the twist-4 parton distribution introduced in  \cite{JaSo, EFP, JWQ}:
\begin{equation} 
   T_1 (x) = \int \frac{ d\lambda}{8\pi} e^{i\lambda x P^+} \langle P\vert  \bar \psi(0) \gamma_\alpha \gamma^+ \gamma_\beta D_\perp^\alpha (0) D_\perp^\beta (\lambda n) \psi(\lambda n)   \vert P \rangle,  
\end{equation}    
as an example to show the problem with parton interpretation. This distribution can be related to the twist-4 distribution $E_q$ defined with two quark fields:
\begin{equation} 
 M^2 E_q (x) = (P^+)^2 \int\frac{d\lambda}{2\pi} e^{-i\lambda x P^+} \langle P \vert \bar \psi (\lambda n) \gamma^- \psi(0) \vert P\rangle,  
\end{equation} 
where $M$ is a scale factor to make $E_q$ dimensionless. If we decompose the quark field  into the plus- and  the minus-component which are defined as:
\begin{equation} 
   \psi^{(+)}(x) = \frac{1}{2} \gamma^- \gamma^+ \psi(x), \quad   \psi^{(-)} (x) = \frac{1}{2} \gamma^+ \gamma^- \psi (x),   
\end{equation} 
we realize that in $E_q$ all quark fields are minus components. 
Using  equation of motion, the minus-component can be expressed with the plus-component combined with gauge fields: 
\begin{eqnarray} 
   \psi^{(-)}(x) = \frac{1}{2}   \int_0^{\infty} d \lambda \biggr [\gamma^+ \biggr ( \gamma_\perp^\mu  D_\mu + i m \biggr )  \psi^{(+)} \biggr ] ( \lambda n +x),
\label{17}    
\end{eqnarray} 
where $m$ is the quark mass. 
It is noted that in this solution we assume as usual the $(-)$-component of $\psi$ to be zero at $x^- =\infty$.
In the light-cone gauge gauge links are unit matrices. 
Using this result, we can derive the relation:
\begin{eqnarray} 
M^2 E_q (x) =  - \frac{2}{x^2} T_1 (x)    + 2\frac{ m}{x}  M e(x) -\frac{m^2}{x^2}  f_q (x), 
\end{eqnarray} 
with $e(x)$ is a twist-3 parton distribution defined as
\begin{eqnarray} 
       M e(x) = P^+ \int\frac{d\lambda}{4\pi} e^{- i\lambda x P^+} \langle P \vert \bar \psi  (\lambda n)  \psi  (0) \vert P\rangle.       
\end{eqnarray}
The above discussed distributions are defined without time-ordering. We define $\tilde E_q$ with time-ordered product: 
\begin{equation} 
 M^2 \tilde E_q (x) = (P^+)^2 \int\frac{d\lambda}{2\pi} e^{-i\lambda x P^+} \langle P \vert   T ( \bar \psi (\lambda n) \gamma^- \psi(0))  \vert P\rangle. 
\end{equation} 
If the time-ordering is irrelevant, we have $\tilde E_q =E_q$. We will check by calculating $E_q(x) $ and $\tilde E_q(x)$  of a single quark state whether they have parton interpretations or not. At the considered 
order, as discussed for $f_q$, the interpretation requires that $E_q(x) $ and $\tilde E_q(x)$ should be zero for $x>1$ and $x<0$.

The tree-level contribution to $E_q(x) $ and $\tilde E_q(x)$ of a single quark is $m^2 \delta(1-x)$.  In the following we will consider contributions with $x\neq 1$. The leading order contribution of $x\neq 1$  to $\tilde E_q(x)$ is from Fig.1a. 
The calculation is similar to that of $f_q$ discussed before. It is straightforward to obtain:
\begin{eqnarray} 
  M^2 \tilde E_q (x) &=& \alpha_s C_F \int\frac{d^2k_\perp}{(2\pi)^2}  \biggr [   2 \theta(x) \theta(1-x) \frac{k_\perp^2(k_\perp^2 +m^2 (x(x-4) +5)) }{(1-x) ( k_\perp^2 + m^2 (1-x)^2 )^2} 
 \nonumber\\ 
    && -  \frac{1}{x^2 (1-x)} \biggr ( \theta (x) -\theta(-x) \biggr ) \biggr ] .   
\end{eqnarray}
Unlike the case with $f_q$, here the integration along the semicircle like the one in Fig.2 is not zero. 
This gives the contribution in the second line in the above. From the result we see that $\tilde E_q$ is not zero for $x>1$ and $x<0$.  
The result has power divergences.   

The contributions to $E_q$ are from Fig.1b, 1c and its complex conjugated one.  The contributions from the last two diagrams have the same problem as that we had in the calculation of $f_q$ of the same diagrams. Using the same method discussed there we obtain at the leading order:  
\begin{eqnarray} 
M^2 E_q (x) &=& \alpha_s C_F \int\frac{d^2k_\perp}{(2\pi)^2}  \biggr [   2 \theta(x) \theta(1-x) \frac{k_\perp^2(k_\perp^2 +m^2 (x(x-4) +5)) }{(1-x) ( k_\perp^2 + m^2 (1-x)^2 )^2} 
 \nonumber\\ 
    && +   \frac{2 }{x^2 (1-x)} \theta(-x)  \biggr ] .   
\end{eqnarray} 
It is not zero for $ x<0$,  and it has a power divergence too. Since $E_q$ is not zero for $x<0$, it has no parton interpretation. Similarly, $\tilde E_q$ also has no parton interpretation because of that it is not zero for $x<0$ and $x>1$.  

\par   
It is noticed that the unphysical contributions with $x>1$ or $x<0$ are power-divergent. In usual one may remove the power divergence in dimensional regularization. However, it is unclear if one can do this here. We will discuss this in detail.  The divergence is related partly to the commutator of field operators, which is nonzero 
as a basic property of quantum field theory.  To see this, we build the difference between $E_q$ and $\tilde E_q$. From the above results we have:
\begin{equation} 
  M^2 E_q(x) - M^2\tilde E_q (x) = \alpha_s C_F \frac{1}{x^2 (1-x)} \int \frac{d^2 k_\perp}{(2\pi)^2}. 
 \label{DEE} 
\end{equation} 
The difference is power divergent.  From the definitions of $E_q$ and $\tilde E_q$, we have the difference:
\begin{equation} 
M^2 E_q (x) - M^2 \tilde E_q (x) =  (P^+)^2 \int\frac{d\lambda}{2\pi} e^{-i\lambda x P^+} \theta (-\lambda)
 \langle P \vert \biggr \{ \bar \psi^{(-)} (\lambda n), \gamma^- \psi^{(-)} (0) \biggr \}  \vert P \rangle,  
\end{equation}  
The difference is related to the commutator of minus-components of quark fields.   
The commutator may be calculated if we use the light-cone quantization, where $x^+$ is taken as the time and QCD is then canonically quantized 
in the light-cone gauge $n\cdot G=0$\cite{KS,ET}.  The canonical equal-time commutation relation for the 
$+$-components of $\psi$ is: 
\begin{equation} 
  \biggr \{ \psi ^{(+)}  (x), \bar \psi^{(+)} (0) \biggr \}  \biggr \vert_{x^+ =0} = \frac{1}{2} \gamma^- \delta (x^-) 
    \delta^2 (x_\perp) .
\label{C25}      
\end{equation} 
It is noted that in the case of $f_q$ the difference  is related 
to the commutator of $+$-components. The commutator is a constant, whose contribution is subtracted
as discussed in \cite{T4Ja,T2PDF}. This is the reason why time-ordering is irrelevant for $f_q$. 
\par 

Using the result in Eq.(\ref{17}) and the commutator in Eq.(\ref{C25}), one can find the commutator of minus-components of quark fields: 
\begin{eqnarray} 
 \biggr \{ \bar \psi(\lambda n), \gamma^- \psi(0) \biggr \}  &=& - \frac{1}{4} g_s^2 \delta^2 (0) 
    \int_0^\infty d\lambda_0  \int_{\lambda}^{\infty} d\lambda_1 \biggr [ 2 \delta (\lambda_1-\lambda_0) G_\mu^a (\lambda_1 n ) G^{a,\mu} ( \lambda_0 n ) 
\nonumber\\    
     &&  + i C_F \epsilon(\lambda_1-\lambda_0) \bar\psi(\lambda_1 n ) \gamma^+ \psi(\lambda_0 n  ) \biggr ] + \cdots, 
\label{E26}     
\end{eqnarray}
where $\cdots$ stand for constant terms, whose contributions are subtracted.  $\epsilon(\lambda)$ is given as $\theta(\lambda)-\theta(-\lambda)$. The $\delta^2(0)$ is the $\delta$-function in the transverse space:
\begin{equation} 
  \delta^2 (x_\perp) = \int\frac{d^2 k_\perp}{(2\pi)^2} e^{ ik_\perp\cdot x_\perp}, \quad 
   \delta^2 (0 ) = \int\frac{d^2 k_\perp}{(2\pi)^2} . 
\end{equation}      
At the considered order,  only the quark part in Eq.(\ref{E26}) contributes. The gluon part contributes at the order of $\alpha_s^2$. 
From the commutator we have for the single quark state: 
 \begin{equation}     
   M^2 E_q(x) - M^2\tilde E_q (x) = \alpha_s  C_F  \frac{1}{x^2 (1-x)} \delta^2 (0), 
\end{equation}   
which is the same as in Eq.(\ref{DEE}).

In $E_q$ and $\tilde E_q$ the transverse momenta of the parton are integrated. One may define their Transverse-Momentum-Dependent(TMD) versions:
\begin{eqnarray} 
  E_q (x, k_\perp ) &=&  (P^+)^2 \int\frac{d\lambda d^2 x_\perp }{ (2\pi)^3 } e^{-i\lambda x P^+ - i x_\perp\cdot k_\perp} \langle P \vert    \bar \psi (\lambda n +x_\perp) \gamma^- \psi(0)  \vert P\rangle, 
\nonumber\\
  \tilde E_q (x, k_\perp ) &=&  (P^+)^2 \int\frac{d\lambda d^2 x_\perp }{ (2\pi)^3 } e^{-i\lambda x P^+ - i x_\perp\cdot k_\perp} \langle P \vert   T ( \bar \psi (\lambda n +x_\perp) \gamma^- \psi(0))  \vert P\rangle. 
\end{eqnarray} 
One can read the results at leading order of these TMD parton distributions from the above results:
\begin{eqnarray}
   \tilde E_q (x, k_\perp) &=&\alpha_s C_F \biggr [  2 \theta(x) \theta(1-x)\frac{k_\perp^2(k_\perp^2 +m^2 (x(x-4) +5)) }{(1-x) ( k_\perp^2 + m^2 (1-x)^2 )^2} 
 - \frac{1}{x^2 (1-x)} \biggr ( \theta (x) -\theta(-x) \biggr ) \biggr ] ,    
\nonumber\\     
E_q (x ,k_\perp)   &=& \alpha_s C_F \biggr [ 2 \theta(x) \theta(1-x)\frac{k_\perp^2(k_\perp^2 +m^2 (x(x-4) +5)) }{(1-x) ( k_\perp^2 + m^2 (1-x)^2 )^2} 
   +  \frac{2}{x^2 (1-x)} \theta(-x) \biggr ], 
\nonumber\\
     E_q(x, k_\perp)  &-&  \tilde E_q (x, k_\perp) = \alpha_s  C_F  \frac{1}{x^2 (1-x)}.       
\end{eqnarray} 
Again, the results are different for the distributions defined with or without time-ordering.

Our next example is the twist-4 gluon distribution $E(x)$   introduced recently in \cite{Ji1}. 
This distribution can be defined without  or with time-ordering: 
\begin{eqnarray} 
 M^2 E(x)  &=&  P^+ \int \frac{d\lambda}{2\pi} e^{-i x \lambda P^+ } \langle P \vert G^{ a,+-} (\lambda n) G^{a,+- } (0) \vert P\rangle,
 \nonumber\\
  M^2 \tilde E(x)  &=&  P^+ \int \frac{d\lambda}{2\pi} e^{-i x \lambda P^+ } \langle P \vert  T \biggr ( G^{a,+-} (\lambda n) G^{a,+- } (0) \biggr )  \vert P\rangle.  
\label{D1} 
\end{eqnarray}  
Although it is  not known whether the distribution can be related to any real physics process,  this distribution is particularly interesting, because of that  it is a special piece of the twist-4 parton distribution defined as 
\begin{equation} 
   F(x) = P^+\int \frac{d\lambda}{2\pi} e^{-i x \lambda P^+ } \langle P \vert G^{ a,\mu\nu} (\lambda n) G^{a}_{\mu\nu } (0) \vert P\rangle.  
\end{equation} 
The first moment of $F$ plays an important role in the decomposition of proton mass studied in \cite{JiMD}.   An one-loop study of $F$ has been performed in \cite{HZ}. In the light-cone gauge, the true dynamical fields are $G_\perp^\mu$.  $G^{-}$ can be expressed with $G^\mu_\perp$. In $F(x)$ there is a part
which only consists of $G^-$-fields. It is the distribution $E(x)$ defined in the above.

\begin{figure}[hbt]
	\begin{center}
		\includegraphics[width=12cm]{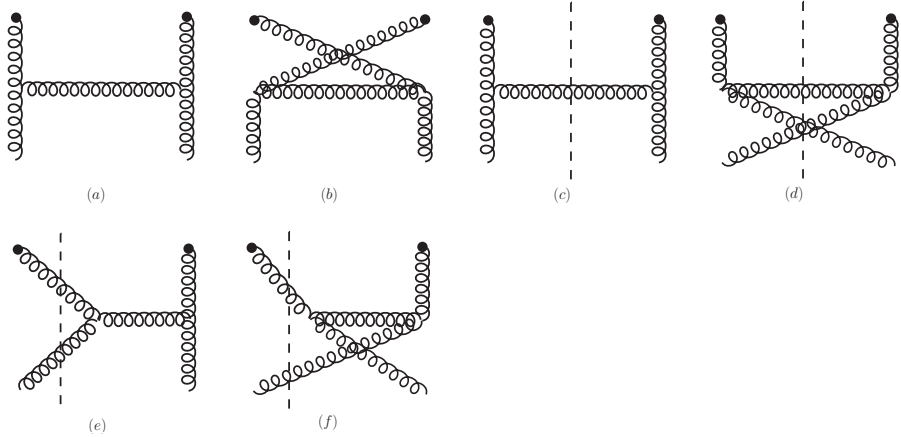}
	\end{center}
	\caption{ Diagrams for  $E(x)$ and $\tilde E(x)$ of  a single gluon state.   }
\end{figure}

At the leading order we calculate $E(x)$ and $\tilde E(x)$  of a gluon state for our purpose. At the considered order contributions are from diagrams given in Fig.3.  $\tilde E (x)$ receives contributions 
from uncut diagrams, i.e., Fig.3a and Fig.3b. The calculation is similar to that of $f_q$ discussed before, 
but with the difference that the integrals along semicircles like the one in Fig.2 are not zero.   
We have at the leading order:
\begin{eqnarray} 
M^2 \tilde E(x) =  \alpha_s N_c \int \frac {d^2 k_\perp}{(2\pi)^2 }  \biggr [   \theta(x) \theta (1-x)  \frac{ x^2 +(1-x)^2 }{1-x}  + ( x\to -x) 
-   \frac{x}{1-x^2 } \biggr ( \theta (x) -\theta (-x) \biggr ) \biggr ].   
\end{eqnarray}              
The distribution is purely power divergent as expected, because of that the initial gluon is real and massless.  We notice that it is not zero for $\vert x\vert >1$. 
Therefore, there is no parton interpretation for this distribution. 

\par 
The cut diagrams in Fig.3 represent the contributions to $E(x)$. The calculations of Fig.3c and Fig.3d 
are straightforward. In the calculations of Fig.3e, Fig.3f and their complex conjugated diagrams, we have the similar 
problem like that in the calculation of $f_q$ from Fig.1c and its complex conjugated diagram. With the method discussed for $f_q$, we can obtain the results from these diagrams. 
Summing all contributions we have the result of $E(x)$ and the difference:
\begin{eqnarray} 
&& M^2 E(x)     = \alpha_s N_c \int \frac {d^2 k_\perp}{(2\pi)^2 }  \biggr [   \biggr (   \theta(x) \theta (1-x)  \frac{ x^2 +(1-x)^2 }{1-x}  + ( x\to -x) \biggr )  
            +  2 \theta (- x)  \frac{x}{1-x^2 } \biggr ] , 
\nonumber\\ 
&& M^2 E(x)  -  M^2 \tilde E(x) =   \alpha_s N_c \frac{x}{1-x^2 }\int \frac {d^2 k_\perp}{(2\pi)^2 }.
\label{EX4}              
\end{eqnarray} 
From our results, we can conclude that there is no parton interpretation 
for the distributions $E(x)$ and $\tilde E (x)$. 

Similarly, we can define TMD parton distributions $E(x,k_\perp)$ and $\tilde E(x,k_\perp)$ by undoing the integration of $k_\perp$ in Eq.(\ref{D1}). The results of the TMD parton distributions are:
\begin{eqnarray} 
 E(x, k_\perp )     &=& \alpha_s N_c  \biggr [   \biggr (   \theta(x) \theta (1-x)  \frac{ x^2 +(1-x)^2 }{1-x}  + ( x\to -x) \biggr )  
            +  2 \theta (- x)  \frac{x}{1-x^2 } \biggr ] , 
\nonumber\\ 
 \tilde E(x, k_\perp) & =&   \alpha_s N_c   \biggr [  \biggr ( \theta(x) \theta (1-x)  \frac{ x^2 +(1-x)^2 }{1-x}  + ( x\to -x) \biggr ) 
-   \frac{x}{1-x^2 } \biggr ( \theta (x) -\theta (-x) \biggr ) \biggr ].   
\end{eqnarray} 
These distributions are finite. Again, they are not zero for $ \vert x \vert >1$. Hence, there is no parton interpretation for these distributions.

\par

The nonzero difference between these two distributions 
indicates that they are  two different distributions. The difference again is determined by the commutator of two $G^{+-}$'s. 
This commutator can be calculated by using light-cone quantization, where the canonical  commutation relation is\cite{KS,ET}: 
\begin{equation}
  \biggr [ G_\perp^{a,\mu} (y), G_\perp^{b,\nu} (0) \biggr ] \biggr\vert_{y^+ =0}  = -\frac{1}{4} i \delta^{\mu\nu} \delta_{ab} \epsilon  ( y^-) \delta^2 (\vec y_\perp).  
 \label{C35} 
\end{equation} 
Using equation of motion one can express $G^{+-}$ with $G_\perp^\mu$-fields.   Doing the similar analysis as done for 
the case of $E_q$ and $\tilde E_q$,  we directly obtain the difference which agrees with that given in Eq.(\ref{EX4}). We also notice that the symmetry of $E(x)=E(-x)$ does not exist because the order 
of two $G^{+-}$ fields in the definition of $E(x)$ can not be changed freely.

\par 
From our discussion at the beginning,  the parton interpretation of the twist-4 quark distribution $E_q$ and $\tilde E_q$ at the order exists only if the distributions are zero for $x<0$ and $x>1$, and that of the twist-4 gluon distributions  $E$ and $\tilde E$  exists only if the distributions are zero for $\vert x \vert >1$.  Our results show that the TMD quark- and gluon distributions have no parton interpretation, because 
of that these distributions are nonzero in the $x$-regions where  the parton interpretation does not exist.    The corresponding parton distributions appearing in collinear factorizations are also nonzero in the $x$-regions, which parton interpretations are not allowed.  
However,  it is subtle in this case to conclude whether parton interpretations exist for the collinear parton distributions at twist-4 or not. 

From our results, the nonzero contributions of the colinear parton distributions in the $x$-regions  which are not allowed by  parton interpretations, are power-divergent. The standard procedure of renormalization with  dimensional regularization is to simply 
set the power-divergent part to be zero or to remove it away. 
Then the parton interpretation can be recovered, at least at the order considered.  
This is different than the case with twist-2 parton distributions, where the parton interpretation is independent to how they are regularized and then renormalized, or the support regions
of the distributions are the same before- and after renormalization. 
But, it may not be correct to simply remove the power-divergent part in the sense of how to correctly 
factorize perturbative effects from nonperturbative ones.  
It is well-known that the perturbative coefficient functions of twist-2 operators still contain soft contributions, which result in the appearance of renormalons.  These contributions are related to the power-divergent parts 
of twist-4 operators because of that the parts dictate the mixing with twist-2 operators.  A consistent separation should be to include these soft contributions into twist-4 parts. Then  
one can not simply neglect the power-divergent parts. 
A detailed discussion about the issue with a twist-4 local operator in DIS can be found in \cite{PJi}.
It is unclear in general how to renormalize the twist-4 parton distributions with power-divergences in the connection to renormalon problems.  
If one uses other regularizations like lattice, the power-divergences can not be ignored. They are renormalized with twist-2 operators or operators of lower dimensions. This can make calculations 
of twist-4 parton distributions with Lattice QCD complicated, e.g., with the method of pseudo-parton distributions in \cite{XJiLa}.

The generic structure of twist-4 parton distributions contain the matrix element of operator products 
like 
\begin{equation} 
   \langle P \vert \phi_1 (\lambda_1 n) \phi_2 (\lambda_2 n) \phi_3(\lambda_3 n) \phi_4 (0)\vert P\rangle,
\label{GT4}     
\end{equation}    
where the field $\phi_i(i=1,2,3,4)$ can be one of the fields $\psi^{(+)}$, $\bar\psi^{(+)}$ or $G_\perp^\mu$.  The order of operators matters. If one changes the order of two fields, e.g., 
changes $\phi_2 (\lambda_2 n) \phi_3(\lambda_3 n)$ to $\phi_3 (\lambda_3 n) \phi_2(\lambda_2 n)$, the difference of the two products is a constant. This results in that the difference for the twist-4 matrix element in Eq.(\ref{GT4}) with the change will be proportional to the matrix element $\langle P \vert \phi_1 (\lambda_1 n)  \phi_4 (0)\vert P\rangle$. In general, such a matrix element is not zero. Therefore, different orders 
give different parton distributions. In the case of twist-3 parton distributions, the involved matrix elements contain three field operators. Changing the order of two field operators will result in 
that the difference between two twist-3 matrix elements is proportional to the matrix element 
of one field operator, which is zero. Therefore, the order plays no role in definitions of twist-3 parton distributions.  

It is noted that the difference between twist-4 parton distributions with different order of operators is power divergent, because of that commutators between two field operators are proportional to $\delta^2 (0)$ as given in Eq.(\ref{C25},\ref{C35}).  If we set the power-divergent part to be zero as usually done with dimensional regularization, the evolution equations of twist-4 parton distributions are DGLAP type integro–differential equations, as explicitly shown at one-loop for distributions defined with non-singlet twist-4 operators in \cite{JiBe}.  It is unclear if such DGLAP type evolution equations exist if the power-divergent contributions are subtracted with counter terms.

In QCD factorization of exclusive processes, instead of various parton distributions, light-cone distribution amplitudes or 
light-cone  wave-functions are involved in theoretical predictions. These amplitudes or wave functions are defined as matrix elements by sandwiching products of operators between a single hadron state and vacuum state. 
They are also classified with twist. With our results it is expected that light-cone distribution amplitudes or 
light-cone wave-functions at twist-4 or beyond will not have a parton interpretation.  

To summarize: We have started to study one-loop calculation of twist-2 quark distribution to explicitly show its parton interpretation, where we have solved the problem of  a possible divergence in the region of $0 > x >-1$. With this solution we have calculated two twist-4 parton distributions in the full $x$-region to show that twist-4 parton distributions have no parton interpretation, because the momentum fraction carried by partons in the twist-4 quark distribution can be larger than one or smaller than 0, while the twist-4 gluon distribution is nonzero for $\vert x\vert >1$. 
The collinear twist-4 parton distributions are purely power-divergent in the $x$-regions where  no parton interpretations exist.  This power divergence may not be simply ignored. Its implication is discussed.

\par\vskip40pt

\noindent
{\bf Acknowledgments}
\par
The work is supported by National Natural Science Foundation of P.R. China(No.12075299,11821505, 11847612,11935017 and 12065024)  and by the Strategic Priority Research Program of Chinese Academy of Sciences, Grant No. XDB34000000.

\par

\end{document}